\documentclass[twocolumn,showpacs,preprintnumbers,amsmath,amssymb]{revtex4}

\usepackage{epsfig}
\usepackage{dcolumn}
\usepackage{bm}

\begin{document}

\title{Local Magnetic Susceptibility of the Positive Muon in the Quasi
1D S=1/2 Antiferromagnet dichlorobis (pyridine) copper (II) }
\author{J.A. Chakhalian, R.F. Kiefl, R. Miller and  J. Brewer}
\affiliation{Department of Physics and Astronomy, UBC,\\
Vancouver, BC V6T 1Z1, Canada }
\author{S.R. Dunsiger and  G. Morris}
\affiliation{Los Alamos National Lab,
 MST-10, MS K764, Los Alamos, NM 87545, USA}
\author{ W.A. MacFarlane}
\affiliation{Chemistry Department, University of British Columbia, Vancouver, 
Canada}
\author{J.E. Sonier}
\affiliation{Department of Physics, Simon Fraser University,\\
Burnaby, BC V5A 1S6, Canada}
\author{S. Eggert}
\affiliation{Institute of Theoretical Physics, Chalmers University of Technology and
G\"{o}teborg University, S412 96 G\"{o}teborg, Sweden}
\author{I. Affleck}
\affiliation{Physics Department, Boston University,
 590 Commonwealth Ave., Boston, MA02215, USA}
 \altaffiliation[Also at ]{Department of Physics and Astronomy, The University of British
Columbia, Vancouver BC, V6T 1Z1, Canada}
\author{A. Keren}
\affiliation{Physics Department, Technion, Israel Institute of
  Technology, Haifa 32000, Israel}
\author{M. Verdaguer}
\affiliation{Laboratoire de Chimie Inorganique et Materiaux
  Moleculaires, Unite CNRS 7071, Universite Pierre et Marie Curie,
  75252 Paris, France}

\date{\today}

\begin{abstract}
  We report muon spin rotation measurements of the local magnetic
  susceptibility around a positive muon in the paramagnetic state of
  the quasi one-dimensional spin 1/2 antiferromagnet dichlorobis
  (pyridine) copper (II).  Signals from three distinct sites are
  resolved and have a temperature dependent frequency shift which is
  significantly different than the magnetic susceptibility.  This
  difference is attributed to a muon induced perturbation of the spin
  1/2 chain. The obtained frequency shifts are compared with Transfer 
  Matrix DMRG numerical simulations. 
\end{abstract}

\maketitle  Novel magnetic effects are predicted for a non-magnetic impurity in
  a one dimensional spin 1/2 antiferromagnetic chain
  \cite{ref:1,ref:2,ref:3}. In particular, at low temperatures
  the magnetic susceptibility in the region of a perturbed link is
  expected to differ dramatically from the uniform bulk susceptibility.
  Furthermore, the effects of such a perturbation propagate far along
  the chain and differ depending on whether the perturbation is link
  or site symmetric.  The effect is closely related to Kondo screening
  of a magnetic impurity in a metal, and arises in part because of the
  gapless spectrum of excitations which characterizes a Heisenberg
  spin 1/2 chain.  Although truly one dimensional spin 1/2 chains have
  no long range ordering above $T=0$, real materials always exhibit 3D
  N$\acute{e}$el ordering due to the finite interchain coupling,
  $J_\perp$. Nevertheless the one dimensional properties can be
  studied down to low temperatures ($T\ll J$) in quasi one dimensional
  systems where $J_\perp \ll J_\parallel$.

A $\mu$SR experiment is an ideal way to test such  ideas since 
the muon acts as both the impurity and the 
probe of the local magnetic susceptibility. We anticipate that the
positively charged muon will distort the crystal lattice, thereby
altering  the exchange coupling between the magnetic ions in
the vicinity of the muon. The resulting modification of the local
susceptibility will  be reflected in the muon frequency shift.

In this paper we report the first muon spin rotation measurements 
on a powdered sample of dichlorobis (pyridine) copper (II)
(CuCl$_2\cdot$2NC$_5$H$_5$) or CPC, which is a well known quasi 1-D
Heisenberg S=1/2 antiferromagnetic salt \cite{ref:4,ref:5}.
We find evidence for three magnetically inequivalent muon sites where
the muon localizes upon thermalization. 
In particular the local spin susceptibility as measured by the muon
frequency shift for two sites displays temperature dependence which is
distinctly different from the bulk magnetic susceptibility.
This effect is attributed to a muon induced perturbation of the local
spin susceptibility.

CPC has a monoclinic crystal structure ($P2_1/n$ space group) and
consists of coplanar units assembled into polymeric chains in which
each Cu$^{2+}$ ion is surrounded by four chlorine anions and two
nitrogen atoms (see Fig. 1). Each Cu$^{2+}$ ion has two
Cl$^-$(1) ions (2.28 \AA) located in the $a-b$ plane and two
more distant Cl$^-$(2) ions (3.05 \AA) located on adjacent planes
in the chain as illustrated  in Fig 1.  The angle between the copper--chlorine and the
copper--nitrogen bonds is close to $90^o$.  The in-chain copper ions
are separated by a distance of 3.57 \AA, compared to the interchain
nearest-neighbor separation of $b=8.59$ \AA \cite{ref:4}.
This large interchain separation assures a high degree of
one-dimensionality.

In order to verify the effect of the $\mu^+$ perturbation and to test
the theory we first measured the bulk susceptibility $\chi$ in fields
of 0.5 T without the perturbing influence of the muon.  The
data were fit to the theory of Eggert, Affleck and Takahashi\cite{ref:5}
and precise values for the interchain coupling $J$ and $g-$factor
were obtained.  The excellent agreement with data is evident from
Fig. 1. Note, although the theoretical procedure was developed
to deal with the impurity problem, the unperturbed case is also an
important test of the theory.  Figure 1 shows the d.c.
susceptibility of CPC along with the best fit curve according to the
theoretical calculation.  Within experimental limits the measured
susceptibility $\chi(T)$ is close to that reported earlier
\cite{ref:6,ref:7} but more accurate.  The
measured bulk susceptibility follows a Curie law at high temperatures,
goes through a maximum around $T=17.8$ K and then the slope starts
increasing again.  As seen in Fig. 1, the theoretical fit to
the experimental data is excellent over the entire temperature range
with deviations of less than 1$\%$.  The best fit yields a value of
the intrachain Heisenberg coupling $J$ of 27.32(30) K and a $g-$factor 
of 2.08(1).
This estimate of $J$ is about 2$\%$ larger than previously reported \cite{ref:6,ref:7}.
This extremely good fit constitutes strong evidence for the validity
of the calculation and manifests a further improvement in theory \cite{ref:5}.

All $\mu$SR measurements were performed at the M20 beamline at TRIUMF
which delivers nearly 100$\%$ spin polarized positive muons with a
mean momentum of 28 MeV/c.  The muon spin polarization was rotated
perpendicular to the axis of the superconducting solenoid and muon
beam direction.  The magnitude of the applied magnetic field $H=0.4$ T
was chosen to provide a good balance between the magnitude of the
frequency shift which increases with field and the amplitude of the
$\mu$SR signal which eventually diminishes with increasing field due
to the finite timing resolution of the detectors.  The transverse
field precession measurements were all performed with a special
cryostat insert which allows spectra to be taken on the sample and on
a reference material simultaneously \cite{ref:8}.
 
Figure 2 shows frequency spectra at 200 K and 8.6 K
 which were obtained by fast Fourier transforming the muon spin
precession signal, which is analogous to the free induction decay in
an NMR experiment.  Near room temperature one observes a single narrow
line, which is attributed to fast muon diffusion whereby the dipolar
interactions with nuclear magnetic moments are motionally averaged.
As the temperature decreases, the line becomes noticeably broadened
and eventually splits into three  frequency lines as the temperature
drops below 25 K (see Fig. 2).  The best least-square fits show that
there are two fast relaxing $\mu$SR signals (labeled as S1 and S2) with
small amplitudes $A_{\rm S1}=0.06$ and $A_{\rm S2}=0.03$
and one slow relaxing signal (S3) with a large amplitude
$A_{\rm S3}=0.15$.  
From this observation, it is clear that muons occupy more than one
magnetically inequivalent sites. Note from the spectrum at 8.6 K in Fig.
2 that three satellite lines are well resolved, implying
three magnetically inequivalent muon sites.  Above 30 K the lines
merge due to the decreasing local spin susceptibility.
 
Because the measurements of the muon precession frequency signal in
the CPC sample and a reference material (silver) were taken
simultaneously, many systematic effects are eliminated. 
After correcting for the temperature independent Knight shift
in Ag (+94 ppm) \cite{ref:9} and the small difference in field
between the reference and sample (22 ppm) we obtain the frequency
shifts for the three sites shown in Fig. 3. 
 A few important observations are in order.  First,
since the experiment was performed on a powdered CPC sample, the
dipolar interaction contributes only to the linewidth and thus the
magnitude of the frequency shift should depend only on the contact
interaction \cite{ref:10}.  The contact hyperfine interaction in CPC is attributed
to either, direct overlap of the wave function tails of the magnetic
electrons with the $\mu^+$, or to the super-transferred hyperfine
field arising from the covalency effects.  Considering the localized
nature of the Cu$^{2+}$ $d$-orbital, the latter effect is more likely.
In this picture, the implanted muon can be viewed as competing for
bonding to the Cl$^-$ ions with some degree of spin density transfer
onto the $\mu^+$ \cite{ref:10_1}.  

In CPC one can identify at least  three inequivalent sites where the muon may localize. 
 Two of them can be associated with a muon interacting with two chlorine ions
 (\textit{i.e.} Cl$^-$--$\mu^+$--Cl$^-$). 
 Note that a similar complex
 has been identified in a variety of ionic solids containing fluorine
 \cite{ref:11} including another well known S=1/2 AF chain KCuF$_3$ \cite{ref:12}. 
 Then the two fast relaxing signals (S1 and S2) can
 be attributed to those two sites where muons effectively `locked'
 between two chloride ions and close to the Cu$^{2+}$ ion. 
The strong temperature dependence of the frequency 
 shifts of the S1 and S2 signals suggests that the muon perturbation
 at these lattice sites is strong.  In contrast, the frequency shift
 of the S3 signal  is relatively small and resembles the bulk magnetic susceptibility
 displaying  a minimum around 14 K which is in the vicinity of the former
 characteristic peak seen in the d.c. susceptibility (see inset in
 Fig. 1). This indicates that the S3 signal is attributed
 to the muons whose influence on the chain is weak.  Considering the
 large interchain distances in CPC (8.59 \AA), one can speculate that
 it is likely that the S3 signal is associated with the muons
 thermalized in the inter-chain space, far from the super-exchange
 path. Also note that the raw frequency shifts of S1 and S2 signals are similar
but have opposite sign. This difference in sign is attributed to site
dependent hyperfine field induced by the polarized Cu$^{2+}$ moments.

This overall muon behavior in CPC is also in agreement with the
theoretical predictions.  The
solid lines in Fig. 3 show a quantitative comparison with the
theoretical calculations\cite{ref:1,ref:2,ref:13,ref:14}
assuming a completely broken link and two completely broken links
, respectively.  Here the muon has been assumed to `feel'
the local magnetic moment of the nearest copper atoms via a contact
interaction of unknown strength.  There are no other adjustable
parameters in this fit.  As seen in Fig.3 the overall agreement is
rather convincing and there is no doubt that the local susceptibility 
of the nearest neighbor Cu is significantly different from the bulk.

 In summary, the local magnetic susceptibility around the muon in
 quasi 1D S=1/2 antiferromagnetic chain compound dichlorobis
 (pyridine) copper (II) (CPC) has been investigated using $\mu$SR
 technique. Signals from three distinct sites are identified and shown
 to have the local magnetic susceptibilities which are different from
 each other and for two locations are also significantly different from the bulk
 susceptibility $\chi$. The theoretical fits capture the effect of
 muon perturbation rather well.  These results confirm the predicted high
 sensitivity of one dimensional spin 1/2 chain compounds to impurity
 effect.

\begin{acknowledgments}
This research was supported by  NSERC, CIAR and the Centre for Materials and
Molecular Research at TRIUMF. The research of Ian Affleck was
supported by the NSF grant DMR-0203159.

\end{acknowledgments}



\begin{figure}
\begin{center}
\epsfig{figure=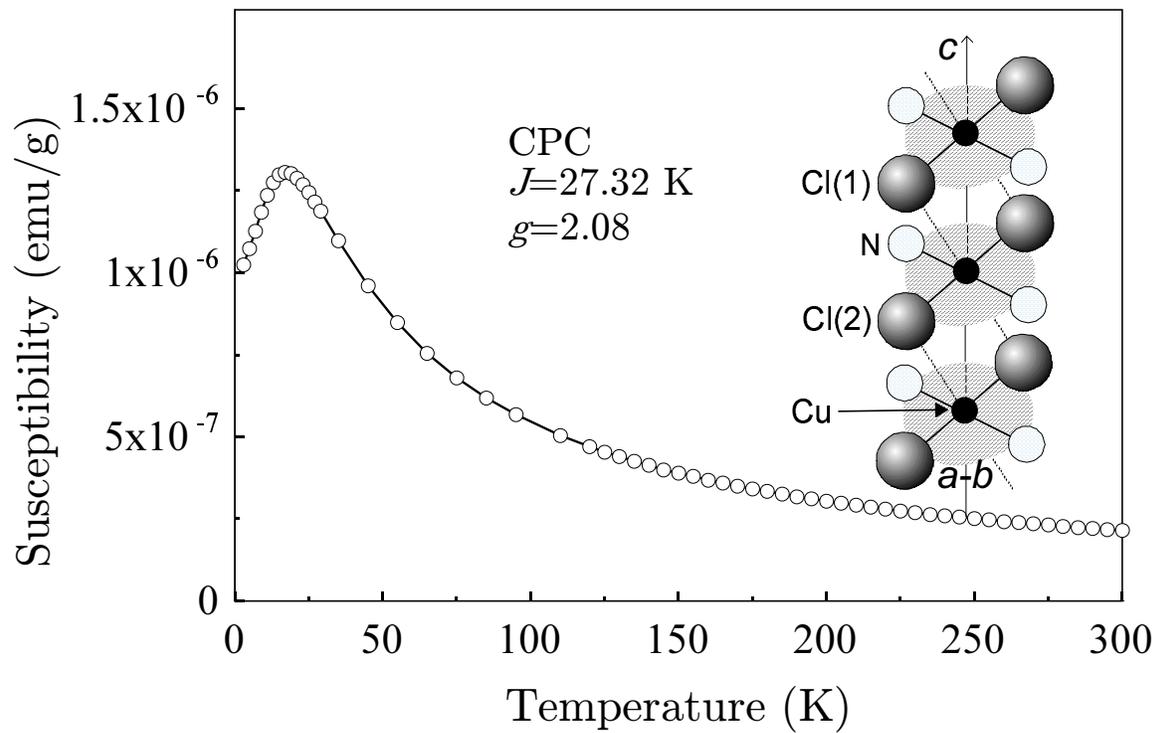,width=17cm, height=!}
\caption{Theoretical fit to the SQUID CPC  data:
The data were taken in an applied magnetic field of 0.5 T. 
The inset show the chain of Cu$^{2+}$ ions (adopted from ref. \cite{ref:6}).}
\label{squid}
\end{center}
\end{figure}

\begin{figure} 
\begin{center}
\epsfig{figure=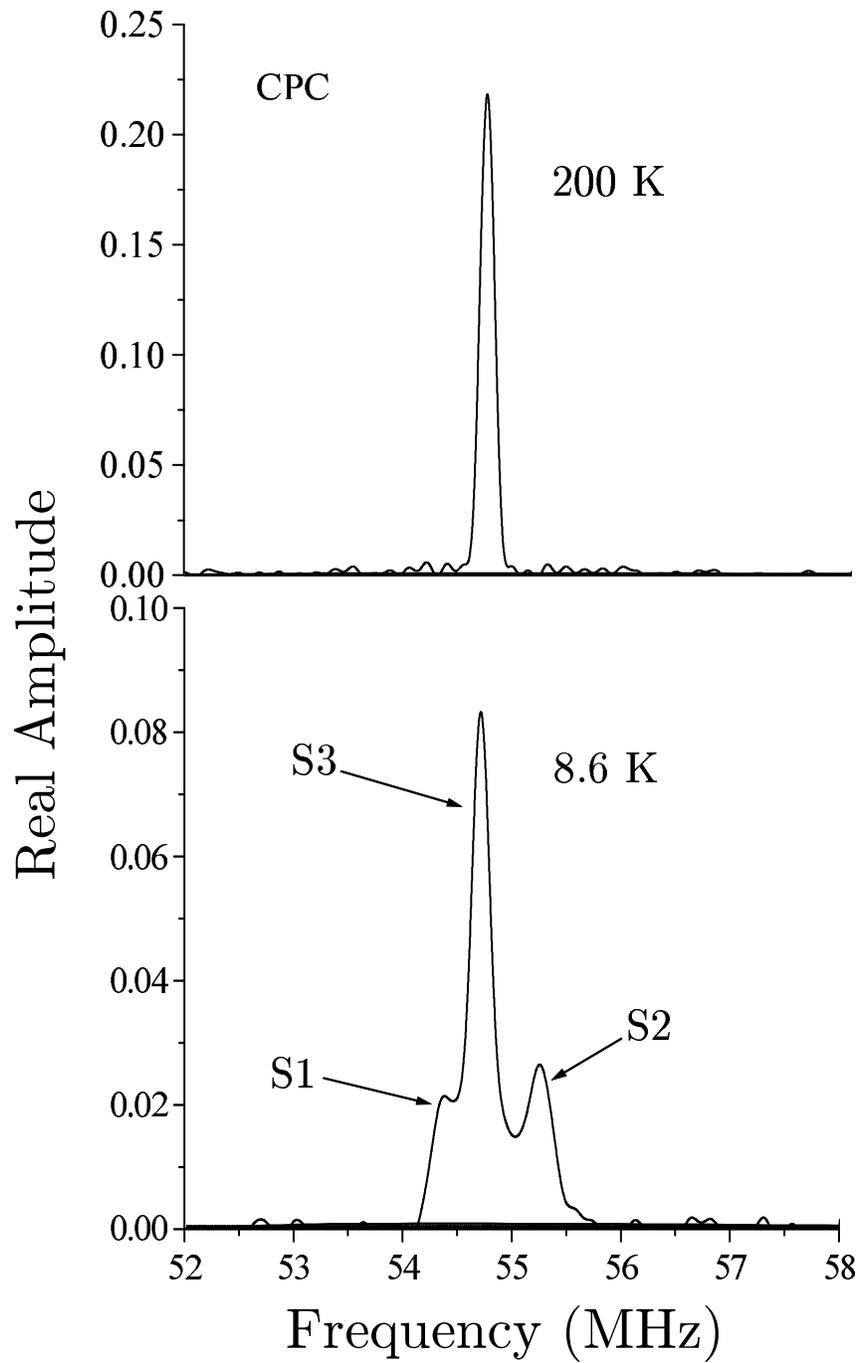,width=15cm, height=!}
\caption{ The evolution of the FFT transforms with temperature in CPC.}
\label{figure2}
\end{center}
\end{figure}

\begin{figure}  
\begin{center}
\epsfig{figure=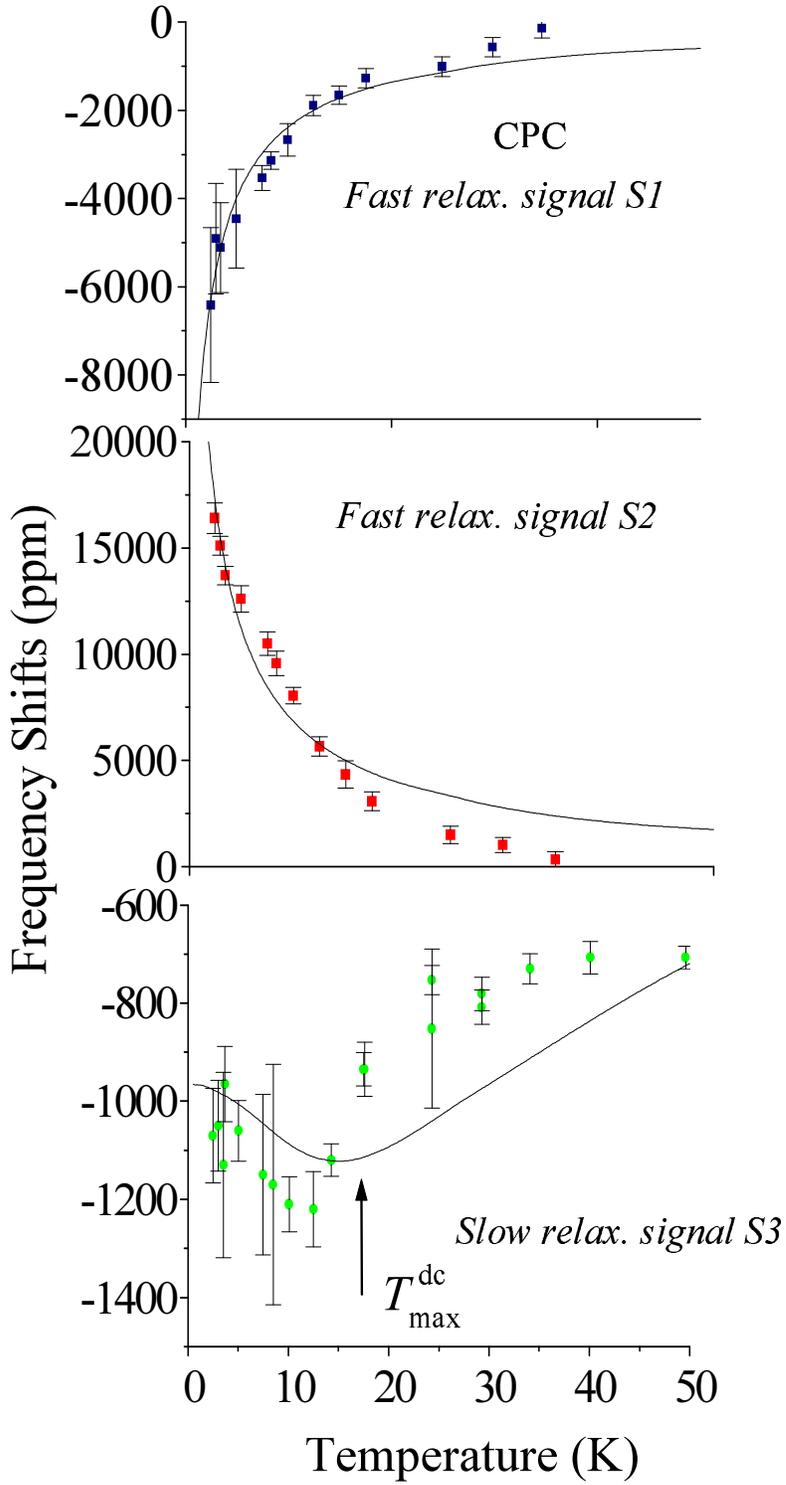,width=15cm,height=!}
\caption{Temperature dependence  of the  frequency shifts  of  the S1,
  S2 and S3 relaxing signals in CPC. In theory simulations, a
  link-symmetric  muon location  is assumed for the signals S1 and S2,
   whereas  a site-symmetric location is assumed for the S3 signal.}
\label{figure5}
\end{center}
\end{figure}

\end{document}